\documentclass[preprint,aps,]{revtex4}

\usepackage{graphicx}
\usepackage{dcolumn}
\usepackage{bm}
\usepackage{multirow}
\usepackage{color}
\usepackage[normalem]{ulem}
\usepackage{amsmath}
\usepackage{gensymb}
\usepackage[colorlinks=true]{hyperref}

\begin{document}

\title{Accessing high optical quality of MoS$_2$ monolayers grown by chemical vapor deposition}

\author{Shivangi Shree$^1$}
\author{Antony George$^2$}
\author{Tibor Lehnert$^3$}
\author{Christof Neumann$^2$}
\author{Meryem Benelajla$^1$}
\author{Cedric Robert$^1$}
\author{Xavier Marie$^1$}
\author{Kenji Watanabe$^4$}
\author{Takashi Taniguchi$^4$}
\author{Ute Kaiser$^3$}
\author{Bernhard Urbaszek$^1$}
\author{Andrey Turchanin$^{2,5}$}

\affiliation{\small$^1$Universit\'e de Toulouse, INSA-CNRS-UPS, LPCNO, 135 Avenue Rangueil, 31077 Toulouse, France}
\affiliation{\small$^2$Friedrich Schiller University Jena, Institute of Physical Chemistry, 07743 Jena, Germany}
\affiliation{\small$^3$Ulm University, Central Facility of Materials Science Electron Microscopy, D-89081 Ulm, Germany}
\affiliation{\small$^4$National Institute for Materials Science, Tsukuba, Ibaraki 305-0044, Japan}
\affiliation{\small$^5$Abbe Centre of Photonics, 07745 Jena, Germany}

\begin{abstract}
Chemical vapor deposition (CVD) allows growing transition metal dichalcogenides (TMDs) over large surface areas on inexpensive substrates. In this work, we correlate the structural quality of CVD grown MoS$_2$ monolayers (MLs) on SiO$_2$/Si wafers studied by high-resolution transmission electron microscopy (HRTEM) with high optical quality revealed in optical emission and absorption from cryogenic to ambient temperatures. We determine a defect concentration of the order of 10$^{13}$~cm$^{-2}$ for our samples with HRTEM. To have access to the intrinsic optical quality of the MLs, we remove the MLs from the SiO$_2$ growth substrate and encapsulate them in hBN flakes with low defect density, to reduce the detrimental impact of dielectric disorder. We show optical transition linewidth of 5~meV at low temperature (T=4~K) for the free excitons in emission and absorption. This is comparable to the best ML samples obtained by mechanical exfoliation of bulk material. The CVD grown MoS$_2$ ML photoluminescence is dominated by free excitons and not defects even at low temperature. High optical quality of the samples is further confirmed by the observation of excited exciton states of the Rydberg series. We optically generate valley coherence and valley polarization in our CVD grown MoS$_2$ layers, showing the possibility for studying spin and valley physics in CVD samples of large surface area.
\end{abstract}


\maketitle

\section{Introduction}
Atomically thin monolayer (ML) transition metal dichalcogenides (TMDs) such as MoS$_2$ and WSe$_2$ are semiconductors with direct bandgaps in the visible to near-infrared region of the optical spectrum \cite{Mak:2010a,Splendiani:2010a,tonndorf2013photoluminescence}. Their strong light matter interaction and fascinating spin and valley properties provide a versatile platform for (opto-)electronics and spintronics \cite{Novoselov:2016a,Mak:2016a,Schaibley:2016a,unuchek2018room,Schneider2018a,Koperski:2017a,dufferwiel2017valley,Scuri:2018a,hong2014ultrafast}. Both fundamental research and potential applications rely on high quality TMD layers to tune optical properties, achieve high emission yields and high carrier mobility in transport. The optical and transport properties do not only depend on the intrinsic quality of the TMD crystal \cite{Amani:2015a,doi:10.1021/acs.nanolett.9b00985}, but also on the surrounding dielectric environment such as the substrate used for growth or deposition \cite{dean2010boron,PhysRevLett.121.247701,PhysRevB.97.201407}. It has been shown in several independent studies that using high quality hexagonal boron nitride (hBN) \cite{Taniguchi:2007a} for encapsulation is crucial to access the intrinsic optical properties of the exfoliated TMD materials \cite{Cadiz:2017a,ajayi2017approaching,wierzbowski2017direct,Jin:2016a,Stier:2018a,2018arXiv181009834M}. For future applications, in addition to high quality the availability of large area TMD MLs is also critical. Individually exfoliated MLs from TMD bulk are typically only tens of micrometers in lateral size, compared to hundreds of micrometers in lateral size and a large number of flakes on the same substrate or even continuous films grown by CVD \cite{lee2012synthesis,tongay2014tuning,doi:10.1021/acsnano.5b01281,eichfeld2015highly,George_2019,yang:2015b,neumann2017opto,antonelou2017mos}. However, so far the optical quality of CVD-grown MoS$_2$ has been considered too low for detailed optical investigations and the vast majority of work especially at low temperature is carried out on individual exfoliated flakes. \\
\indent Here we show that conventional CVD yields MoS$_2$ monolayers with high optical quality. We determine a defect concentration of 10$^{13}$ cm$^{-2}$ for our samples by high-resolution transmission electron microscopy (HRTEM), comparable to defect concentrations for exfoliated MLs from naturally occurring bulk \cite{zhou2013intrinsic,hong2015exploring,vancso2016intrinsic,rhodes2019disorder}. Based on this measurement we therefore expect similar optical quality to exfoliated material. However, our photoluminescence (PL) and reflectivity experiments on as-grown CVD MoS$_2$ MLs showed optical transitions with large inhomogeneous broadening ($\approx$ 50~meV at T=4~K), as commonly reported in the literature for this system  \cite{PhysRevMaterials.2.064003,plechinger2014direct,yu2016engineering}. 
To avoid detrimental effects from ML-substrate interaction and to minimize therefore the impact of disorder \cite{rhodes2019disorder,Archana2019a} we removed the MLs from the SiO$_2$ growth substrate. We subsequently encapsulated the MoS$_2$ in high quality hBN \cite{Taniguchi:2007a} for optical absorption and emission experiments for temperatures between 4~K to 300~K. These hBN encapsulated CVD-grown MoS$_2$ MLs show very narrow optical transition linewidth with 5~meV FWHM at T=4~K, similar to values for exfoliated material \cite{roch2019spin, wierzbowski2017direct}. The PL is dominated by free excitons and not by defects even at cryogenic temperatures. High optical quality of the samples is further confirmed by the observation of excited states of the Rydberg series in absorption for the A-exciton \cite{Robert:2018a,2019arXiv190403238G}. Using above bandgap, polarized laser excitation, we optically generate large valley coherence and valley polarization in our CVD grown MoS$_2$ layers \cite{Jones:2013a}. It is therefore possible to explore spin and valley physics in these high quality CVD samples of large surface area in more detail.\\
\indent The paper is organized as follows, in Section~\ref{crystal} we analyze the crystalline quality of our CVD grown MoS$_2$ monolayers, in Section~\ref{optics} we present optical spectroscopy results which compare emission and absorption of the CVD-grown samples with exfoliated layers, in Section~\ref{sum} we compare with other approaches on TMD growth and encapsulation techniques and summarize our work. More details on experimental techniques and results are given in the supporting information (SI). 

\begin{figure*}
\includegraphics[width=0.95\linewidth]{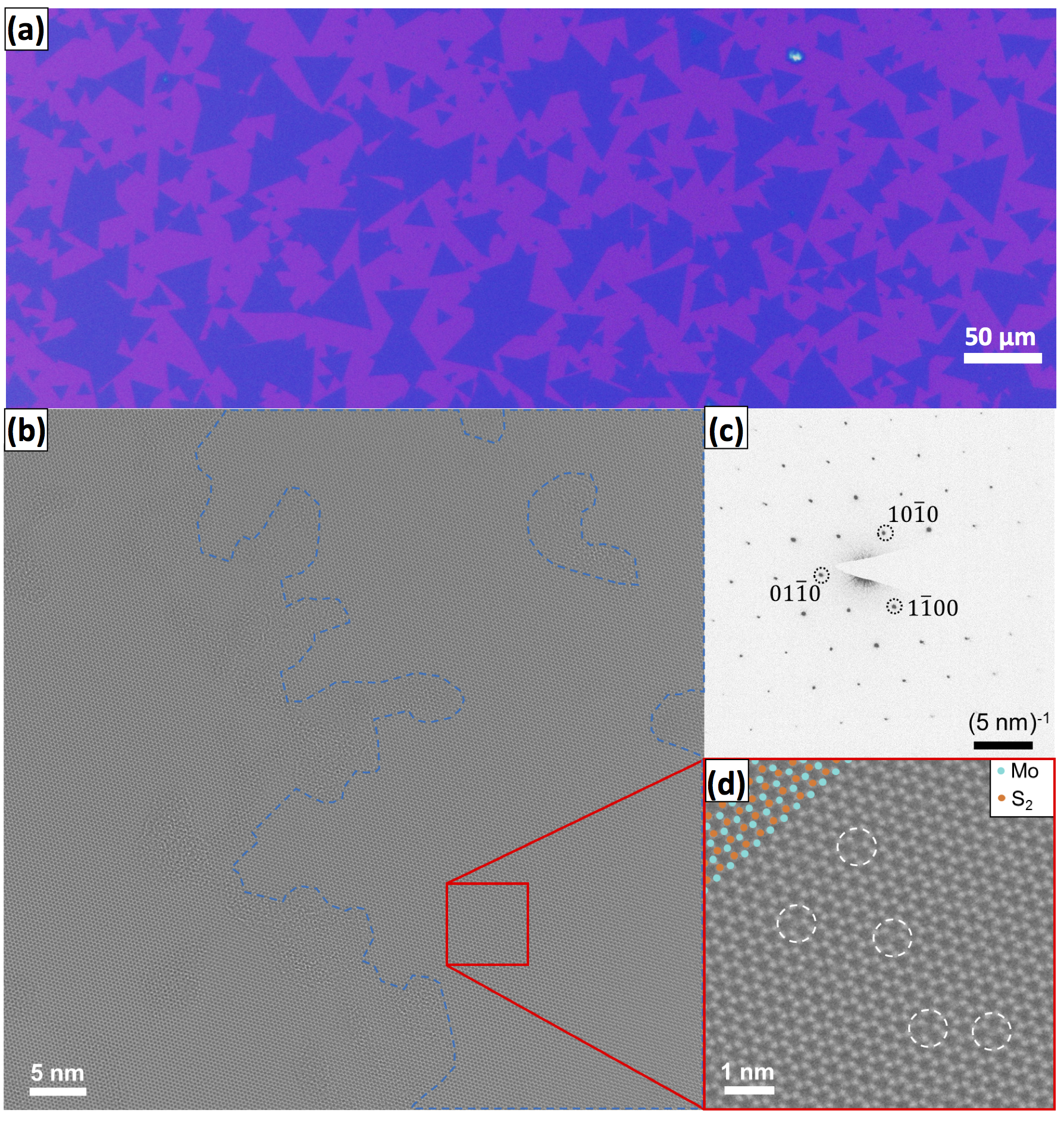}
\caption{\label{fig:fig1} (a) Optical microscope image of as grown CVD MoS$_2$ monolayers on SiO$_2$/Si wafers (300 nm of SiO$_2$) showing their characteristic triangular shapes. (b) 60 kV HRTEM image of MoS$_2$. Defect densities were evaluated by direct counting of vacancies on clean areas as can be seen within the blue framed area. The diffraction pattern in (c) show a high crystallinity over the whole imaged area in (b). The red square in (b) is magnified in (d). For better visualization, different atoms are coloured. Orange dots indicate two stacked-up S atoms (S$_2$) and turquoise solid dots correspond to Mo atoms. White circles mark the positions of vacancies.}
\end{figure*}

\section{Crystal quality from atomic resolution microscopy and optimized growth}
\label{crystal}
The defect density reported in the literature for MoS$_2$ is typically around 10$^{13}$ cm$^{-2}$. The dominant type of point defect varies depending on the material source, as reviewed recently \cite{rhodes2019disorder}: in naturally occuring MoS$_2$ and CVD-grown films, sulfur vacancies can dominate \cite{vancso2016intrinsic}, whereas in synthetic crystals grown by chemical vapour transport (CVT), metal vacancies and antisites tend to dominate \cite{doi:10.1021/acs.nanolett.9b00985}. First, we describe how we quantitatively determine the defect density (i.e. of the dominating sulfur vacancies) in our CVD grown monolayers and compare these data with measurements in the same set-up on exfoliated monolayer. \\
\indent The MoS$_2$ crystals were grown by a modified CVD process in which a Knudsen-type effusion cell is used for the delivery of sulfur precursor, for details of the growth see \cite{George_2019}. Fig. \ref{fig:fig1}a presents an optical microscope image of a typical sample with a high density of ML MoS$_2$ crystals with lateral dimension of several tens of micrometers, the same kind of sample grown under the same conditions in the same set-up investigated in optical spectroscopy experiments in the next Section \ref{optics}. To study the crystalline quality of the samples on an atomic scale we apply high-resolution transmission electron microscopy (HRTEM) with the C$_c$/C$_s$-corrected sub-Angstrom low-voltage electron microscope (SALVE)  operating at 60 kV. In our set-up we carefully optimize high resolution with respect to minimal radiation damage \cite{doi:10.1063/1.4973809}. As can be seen in Fig. \ref{fig:fig1}b-d, the CVD-grown MoS$_2$ MLs show an overall excellent crystallinity with a low defect density. Fig.~\ref{fig:fig1}d shows an enlarged image of the area marked with a red square in Fig. \ref{fig:fig1}b, where some double sulfur vacancies typically present in the ML MoS$_2$ samples \cite{George_2019,doi:10.1063/1.4973809,doi:10.1063/1.4830036,PhysRevLett.109.035503}, can be recognized.  \\
\indent The double sulfur vacancies were directly counted from the HRTEM images as in Fig.~\ref{fig:fig1}d (see also  Fig. S1) to determine a quantitative value for their density $C=V/A$, where $V$ is the number of the sulfur vacancies per area A. In the CVD-grown samples we find a density of the vacancies equal to $7.9(6)\times 10^{13} V/cm^{2}$. The value in the brackets gives the confidence intervals, see  corresponding calculation in the SI. For comparison, we investigated  ML samples prepared by exfoliation from bulk MoS$_2$ crystals (see SI for details). We found a vacancy density of $3.6(6)\times 10^{13} V/cm^{2}$. The measured defect density of CVD-grown ML MoS$_2$ is therefore roughly of the same order of magnitude as for exfoliated MLs from natural or chemical vapor transport (CVT) grown bulk \cite{doi:10.1021/nn500044q,chuang2016low,doi:10.1021/acs.nanolett.9b00985}. Thus based on this HRTEM investigation we conclude that the CVD-grown ML MoS$_2$ possess comparatively high structural quality and therefore high optical quality in emission and absorption experiment of these samples could be expected.
\section{Optical Spectroscopy of CVD-grown MoS$_2$ monolayers}
\label{optics}
\begin{figure*}
\includegraphics[width=0.95\linewidth]{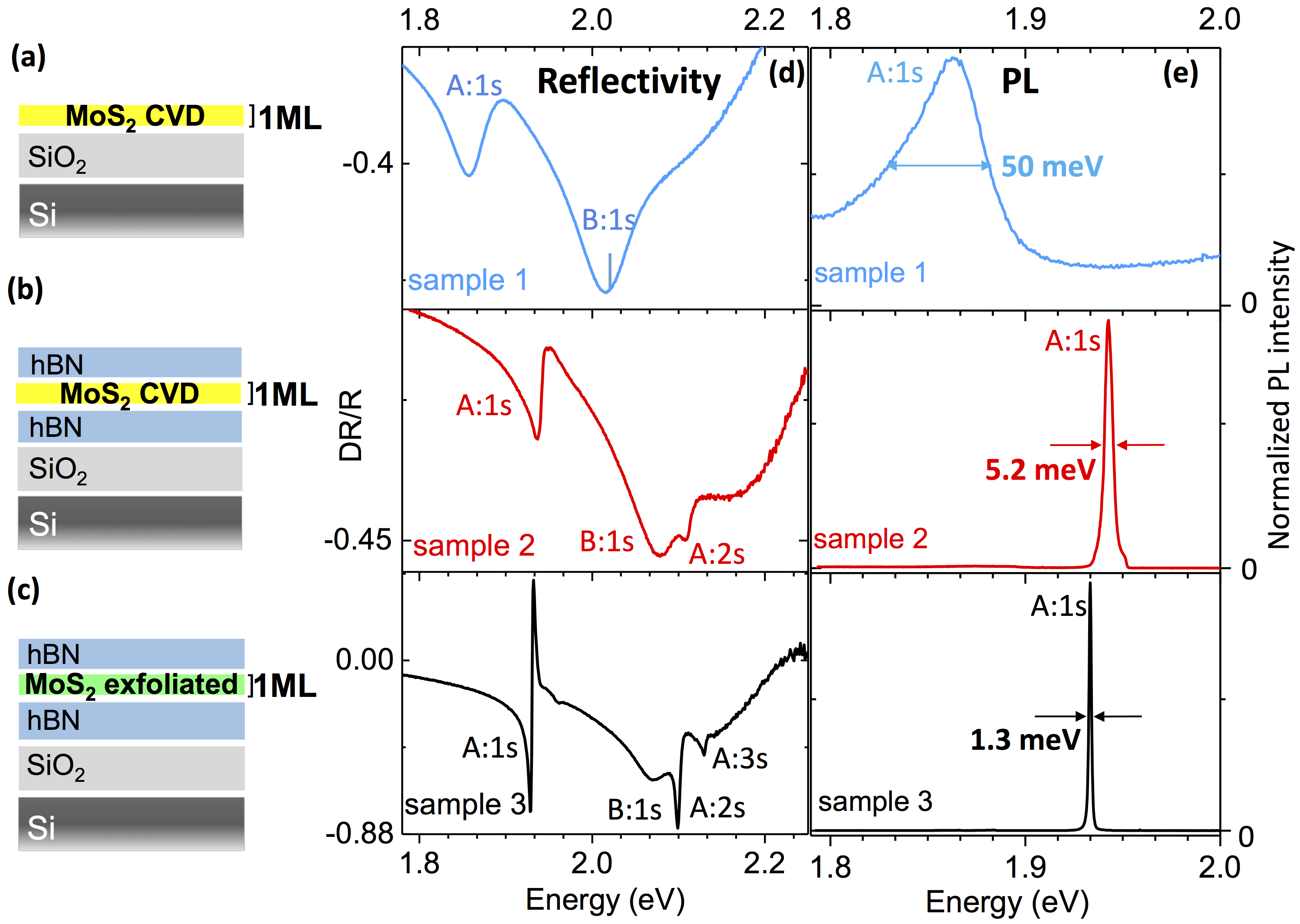}
\caption{\label{fig:fig2} (a) Schematics of Sample 1, as-grown CVD MoS$_2$ ML on SiO$_2$ (b), (c) CVD-grown and exfoliated MoS$_2$ monolayer encapsulated in hBN layers -sample 2 and sample 3, respectively. (d) Differential reflectivity measurements at T=4~K performed on different samples 1,2 and 3. A-exciton and B-exciton transitions are marked for sample 1. For samples 2 and 3 excited state of the A-exciton at higher energies are clearly visible marked A:2$s$, A:3$s$. (e) PL spectra of the as-grown and encapsulated MoS$_2$ in h-BN, respectively at T= 4~K, which highlights the very different linewidth and shift in emission energy. }
\end{figure*}

The optical transition linewidth of ML TMDs contains homogeneous and inhomogeneous contributions. As the radiative broadening is of the order of 1~meV any substantially larger linewidth at low temperature is dominated by inhomogeneous contributions due to imperfections in the ML or the direct environment \cite{Moody:2015a,2018arXiv181009834M}. We study the optical quality in PL and differential reflectivity experiments, using a detection and excitation spot diameter of the order of 1~$\mu$m, see SI for experimental details.  We have performed measurements on MoS$_2$ MLs in three different structures, see Fig.~\ref{fig:fig2}a,b and c. Large-area, as-grown MoS$_2$ films on SiO$_2$ by CVD are represented by sample 1 in Fig.~\ref{fig:fig2}a. To fabricate sample 2, we proceed as follows : First we deposit an hBN flake about 100~nm thick on a SiO$_2$/Si \textit{target} substrate (different from the \textit{growth} substrate).  Then we remove the CVD-grown MoS$_2$ from the growth substrate using a dry stamp \cite{Gomez:2014a} and deposit this layer on top of the prepared hBN flake on the \textit{target} substrate. Finally the MoS$_2$ layer is  covered by a large hBN flake about 10~nm thick. Sample 2 has therefore the following structure: SiO$_2$/hBN/MoS$_2$~ML~CVD/hBN, see figure S4 for an optical contrast image of this van der Waals (vdW) heterostructure. Sample 3 is an exfoliated MoS$_2$ monolayer from commercial bulk MoS$_2$ (from 2D Semiconductors) encapsulated in hBN, similar to exfoliated and encapsulated samples we studied previously \cite{Cadiz:2017a,Robert:2018a}. The difference between sample 2 and 3 is just the source of the MoS$_2$ ML, CVD-grown and exfoliated from commercial bulk, respectively.

First, we discuss differential white light reflectivity measurements at low temperature, $T = 4$~ K. In reflectivity, we typically probe transitions with high oscillator strength. This technique is less sensitive to defect related optical transitions, for this purpose we use PL discussed below. We measure differential reflectivity $(R_{ML}-R_{sub})/R_{sub}$, where $R_{ML}$ is the intensity reflection coefficient of the sample with the MoS$_2$ layer and $R_{sub}$ is the reflection coefficient of the same structure, but without the ML. Note that the overall shape and amplitude of the differential reflectivity signal also depends on cavity effects (thin-layer interference) given by top and bottom hBN and SiO$_2$ layer thickness (see \cite{Robert:2018a} for details).\\
\indent We compare the reflectivity spectra of the CVD-grown films to those of exfoliated MoS$_2$ in hBN, which represent the current state of the art in terms of linewidth close to the homogenous limit, see three panels of Fig.~\ref{fig:fig2}d. In differential reflectivity, the A- and B-exciton transitions are very broad (several tens of meV) in as-grown CVD monolayers (labeled A:1$s$, B:1$s$) for sample 1. However, for sample 2 the A:1$s$ transition is considerably narrower. Near the B-exciton transition energy two different transitions can be distinguished, namely the B:1$s$, much narrower than for sample 1 and in addition the excited state of the A-exciton, namely A:2$s$ \cite{Chernikov:2014a,He:2014a}. The improvement in optical quality of CVD grown MoS$_2$ in hBN is significant compared to the as-grown monolayer from the same sample. Globally the reflectivity spectrum of sample 2 and sample 3 are very similar, showing the high optical quality of our CVD grown samples when they are encapsulated. Seeing the A:2$s$ state in sample 2 is a sign of high optical quality \cite{2019arXiv190403238G} as with the larger Bohr radius we sample a bigger sample volume \cite{Stier:2018a}. For enhanced visibility the A:2$s$ state also the clean dielectric environment i.e. the hBN material with very few defects is important, as excited exciton states are broadened considerably by dielectric disorder \cite{Archana2019a} and are therefore spectrally narrower in encapsulated samples \cite{Manca:2017a}. The A-exciton transition A:1$s$ is redshifted by about 50~meV in sample 1 compared to both samples 2 and 3, which might be due to strain induced in the layer during the cool-down process after growth \cite{plechinger2014direct}. \\
\indent Now we discuss PL spectroscopy of the three samples, a technique considerably more impacted by the presence of defects compared to reflectivity. The PL spectrum of the as-grown MoS$_2$ monolayer on SiO$_2$ is shown as the blue curve in Fig.~\ref{fig:fig2}e using a laser energy of 2.32~eV well above the emission energy. It shows a roughly 50~meV wide A-exciton emission at 1.863~eV, very similar to other spectra reported for as-grown samples in the literature \cite{neumann2017opto}. We also note that the transition energy of neutral exciton A:1$s$ is redshifted compared to standard MoS$_2$ ML both in PL as observed also in reflectivity in Fig.~\ref{fig:fig2}d. The broad linewidth of the PL emission, similar to the broad linewidth measured in differential reflectivity, indicates large inhomogeneous broadening of the A-exciton transition. This is surprising when taking into account the high structural quality measured for this CVD grown material, see  Fig.~\ref{fig:fig1} and hints at detrimental effects from the environment to be at the origin of this inhomogeneous broadening \cite{2018arXiv181009834M}. \\
\indent For sample 2 the PL emission linewidth is clearly reduced by a factor of 10 after encapsulating the CVD-grown MoS$_2$ monolayer in hBN. We also noticed a significant enhancement in PL emission intensity at the A:1$s$ energy after encapsulation as compared to low energy defect related emission, see discussion of Fig.~\ref{fig:fig3} below. The PL linewidth for sample 2 with 5~meV in Fig.~\ref{fig:fig2}e is a typical linewidth reported also for MoS$_2$ exfoliated and encapsulated material at T = 4~K \cite{roch2019spin,wierzbowski2017direct}. So finally these spectra indicate that the comparable structural quality of our CVD layers to exfoliated material also results in comparable optical quality, once the detrimental impact of the substrate (surface roughness and charge fluctuations) are suppressed thanks to the hBN buffers. Sample 3 shows an even narrower PL emission linewidth with 1.3~meV and indicates that further reduction of inhomogeneous broadening for the CVD sample can be targeted in the future by tailoring growth and improving the encapsulation procedure, see discussion in the next section. 

\begin{figure*}
\includegraphics[width=0.95\linewidth]{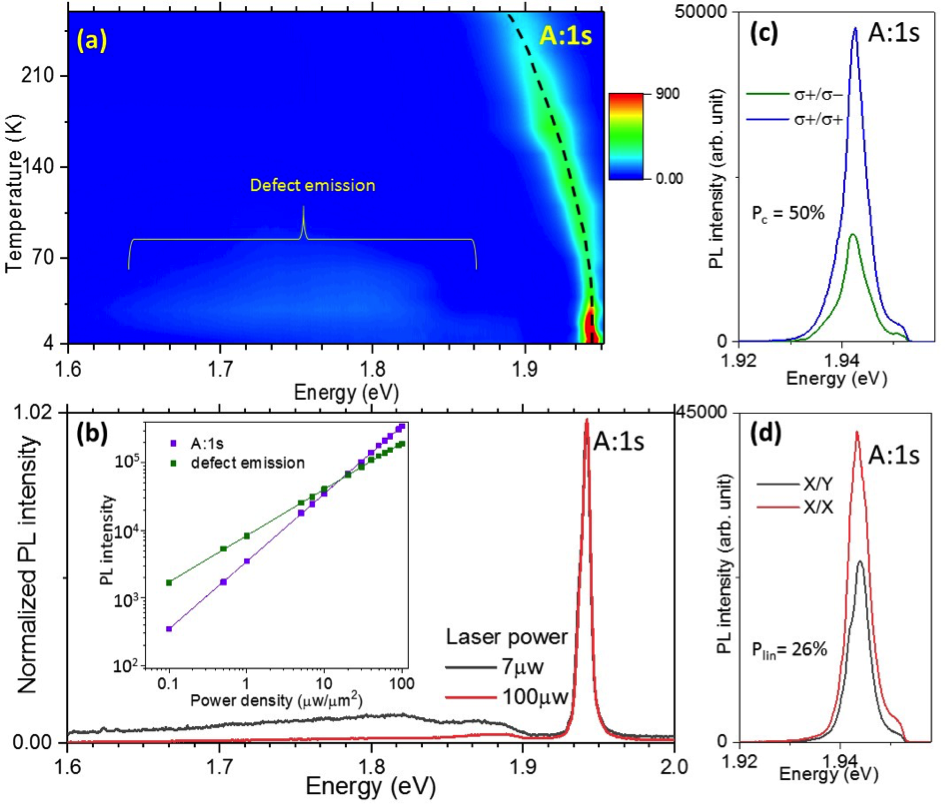}
\caption{\label{fig:fig3} Sample 2. (a) A color contour plot of the temperature dependence of the photoluminescence intensity using a laser excitation density 5~$\mu$W/$\mu$m$^2$. The dotted line tracks the evolution of the peak position of the neutral exciton (optical bandgap A:1$s$) with temperature according to Eq.(1). (b) PL spectra T= 4~K with cw laser at 2.32~eV for excitation power 7~$\mu$W and 100~$\mu$W. Low energy, spectrally broad defect emission is clearly visible at low excitation power, but not at high power. The inset shows a double logarithmic plot of the integrated PL intensity as a function of excitation power for A:1$s$ (purple filled squares) and defect emission (green filled squares). The solid curves represent $I\propto P^{\alpha}$, a fit to PL intensity, yielding an $\alpha$ of 1.0 for A:1$s$ and 0.7 for the defect state. (c) and (d) Polarization-resolved PL at T=4~K following circularly and linearly polarized laser excitation at 2.32~eV, exhibiting efficient exciton valley polarization and valley coherence, respectively. }
\end{figure*}

We now study the emission properties of sample 2 in more detail in Fig.~\ref{fig:fig3}. Figure~\ref{fig:fig3}a shows the PL spectrum for selected temperatures from 260 K to 4 K. The PL spectrum at 4~K exhibits a neutral exciton (A:1$s$) peak centered at 1.944~eV. The linewidth of the A:1$s$ peak is around 5~meV and moreover the PL intensity of the defect related peaks at lower energy is significantly smaller than that of the A:1$s$ peak, we see no clear indication of a charged exciton peak (trion) \cite{Mak:2013a,roch2018quantum}, indicating (close to) charge neutrality. The fact that the PL spectrum is dominated by the main neutral exciton emission is a clear indication of a low defect density, in agreement with the HRTEM results shown in Fig.~\ref{fig:fig1}.  It is very important to note that PL spectra with spectrally narrow A:1$s$ emission and very weak defect emission are observed throughout the sample and not just for a specific detection spot location, see Figure S4.\\
\indent In order to further study the main exciton transition, we measure the emission energy and linewidth as a function of temperature. Figure~\ref{fig:fig3}a shows the A:1s peak position plotted as a function of temperature. The A:1$s$ transition energy can be fitted by the following standard expression for the temperature dependence of a semiconductor bandgap \cite{o1991temperature}:
\begin{equation}
E_G(T)=E_G(0) - S \langle  \hbar w \rangle [\coth (\langle \hbar w\rangle/(2k_BT)  ) -1]
\label{eq1}
\end{equation}
where $E_G(0)$ is the optical band gap at T=0~K, $S$ is a dimensionless electron phonon coupling constant, $k_B$ is Boltzmann’s constant, and $\langle  \hbar w \rangle$ is an average phonon energy. Fitting the data in Fig.~\ref{fig:fig3}a with Eq.~\ref{eq1} yields $E_G(0)$=1.944~eV, S = $1.9 \pm 0.1$ and $\langle  \hbar w \rangle = 20 \pm 2$~meV. These parameters are very close to the values obtained for high quality exfoliated samples, which confirms that the narrow peak of 5~meV observed for sample 2 is indeed the main A-exciton transition (optical bandgap) of the MoS$_2$ monolayer \cite{Cadiz:2017a}. This is further confirmed by the measurement of the same transition energy in PL and reflectivity.\\
\indent The evolution of the linewidth with temperature can be phenomenologically approximated by a phonon-induced broadening \cite{Selig:2016a,Dey:2016a} :
\begin{equation} \gamma= \gamma_0 + c_1 T +  \frac{c_2}{e^{\langle \hbar w\rangle/k_BT}-1} \label{eq2} \end{equation}
where $\gamma_0=5.6 \pm 0.2$~meV,  $c_1=56\pm 6~ \mu$eV/$\mbox{K}^{-1}$ describes the linear increase due to acoustic phonons, and $c_2=31 \pm 2$~meV is a measure of the strength of the exciton-phonon-coupling, respectively. $\langle \hbar w\rangle=20$~meV is the \textit{averaged} energy of the relevant phonons, that we obtained by fitting the optical bandgap energy shift with Eq.(\ref{eq1}) for consistency. Again these extracted values are very similar to reports on exfoliated MoS$_2$ MLs, indicating similar strength of the exciton-phonon interaction in our CVD-grown layers and hence no major additional broadening mechanisms \cite{Cadiz:2017a,Selig:2016a,Dey:2016a,Moody:2015a}. The temperature dependent PL data also shows the low energy ($< 1.9$~eV) defect associated emission disappears above T=150~K. \\
\indent Another indication for high optical quality comes from power dependent PL. The inset of Fig.~\ref{fig:fig3}b displays the spectrally integrated PL intensity $I$, defined as the peak area, as a function of excitation power $P$ (with $I \propto P^\alpha)$ for A:1$s$ (filled purple squares) and  defect states  (filled green squares). As laser power increases, more carriers are generated that fill up gradually all available defect sites, leading to a gradual saturation of the defect assisted emission. This leads to a sublinear behavior when defect emission is plotted as a function of laser power for the defect peak reaching $\alpha= 0.68$. The power dependence of defect emission follows a nonlinear curve, while A:1$s$ is linear with fitted $\alpha \approx 1$. Only at higher excitation powers when exciton-exciton scattering processes become important also the neutral exciton emission will saturate with power, for example due to exciton-exciton annihilation \cite{Kumar:2014b,Mouri:2014a}, but this is beyond the laser power range investigated here. \\
\indent The high quality sample 2 is further confirmed in studies using polarized light in excitation and detection. Studies of optically controlled valley polarization and valley coherence are one of the main motivations for the quest of cleaner TMD materials with narrower linewidth \cite{Schaibley:2016a}. We first investigate exciton spin-valley polarization based on the chiral optical selection rules \cite{Cao:2012a,Xiao:2012a}. \\
\indent In our experiment the MoS$_2$ ML is excited by a circularly polarized ($\sigma^+$) continuous wave He-Ne laser (1.96~eV) with power density 5$\mu$W/$\mu m^2$. We define the circular PL polarization degree as $P_c=\frac{I_{\sigma+}-I_{\sigma-}}{I_{\sigma+}+I_{\sigma-}}$, where $I_{\sigma+}(I_{\sigma-})$ are the intensity of the right ($\sigma^+$) and left ($\sigma^-$) circularly polarized emission. The circular polarization in time integrated experiments depends on the exact ratio of emission time $\tau_{PL}$ versus depolarization time $\tau_{depol}$ as $P_c=1/(1+\tau_{PL}/\tau_{depol})$. We find $P_c \approx 50\%$ for the emission in Fig.~\ref{fig:fig3}c, comparable to previously reported values in exfoliated MoS$_2$ ML samples \cite{Cadiz:2017a,Zeng:2012a,Mak:2012a,Kioseoglou:2012a}. \\
\indent A coherent superposition of valley states, valley coherence, can be generated using linearly polarized excitation \cite{Jones:2013a,Hao:2015a,Wang:2016b}. The MoS$_2$ ML is excited by a linearly polarized (X) laser and we detect the emitted light in both polarization directions X and Y. The linear PL polarization degree $P_{lin}$ is defined as $P_{lin} = \frac{I_X-I_Y}{I_X+I_Y}$, where $I_X (I_Y)$ denotes the intensity of the X and Y linearly polarized emission, respectively. In our experiment we find sizeable $P_{lin}$ of about 26$\%$ for the emission in Fig.~\ref{fig:fig3}d. Please note that the exact values of $P_c$ and $P_{lin}$ critically depend on the laser excitation energy \cite{Wang:2015b,kioseoglou2016optical}, which was not varied here. 
The results in Fig.~\ref{fig:fig3}c and d show that we can generate exciton valley polarization and valley coherence optically, making these large area CVD-grown films highly useful for future the studies of valley and spin physics in monolayer MoS$_2$. Optical generation of spin-valley polarization in TMDs is interesting for hybrid devices that rely on subsequent spin transfer to graphene for effective spin transport \cite{luo2017opto,avsar2017optospintronics}. 

\section{Discussion and Summary}
\label{sum}
The first optical spectroscopy reports for emission in ML MoS$_2$ exfoliated on SiO$_2$/Si substrates have given PL linewidth of the order of 50~meV, even at low temperature \cite{Mak:2010a,Splendiani:2010a,Korn:2011a,Kioseoglou:2012a,Sallen:2012a}. Now in exfoliated and hBN encapsulated samples we approach the radiatively broadened, homogeneous limit with linewidth of 1~meV \cite{Cadiz:2017a,ajayi2017approaching,wierzbowski2017direct,2019arXiv190205036R,2019arXiv190200670F}. This research fields has made impressive progress as the impact of disorder in the ML environment on the measured optical properties of the ML is better understood \cite{rhodes2019disorder,Archana2019a}. In this work we have shown that CVD-grown monolayers, picked up from the growth substrate and encapsulated in exfoliated hBN flakes show high optical quality and allow detailed spectroscopic and valley polarization studies. An important aspect of our work is that the initial growth is performed on SiO$_2$, a commonly available substrate material of industrial grade. 

Other hybrid approaches with CVD growth and exfoliated flakes have also shown promising results using different substrates. In general, the detrimental effects of the substrate material can be studied for monolayers that are lifted of the substrate after growth, for instance, Yu et al \cite{yu2016engineering} report strongly enhanced PL emission for suspended MoS$_2$ and WS$_2$ monolayers. Recent experiments on MoSe$_2$ monolayers encapsulated in hBN and suspended over a micro trench do not only show extremely narrow linewidth close to the homogeneous limit, but also display very little variation of the emission energy in a spatial scan of the suspended region \cite{2019arXiv190108500Z}. Better optical quality for CVD-grown WS$_2$ removed from the substrate has also been reported by Hoshi et al \cite{PhysRevMaterials.2.064003} and for CVD-grown WSe$_2$ by Delhomme et al \cite{doi:10.1063/1.5095573}, confirming this trend. 

Direct growth not on SiO$_2$ but directly on hBN is another way to achieve high quality material, demonstrated for both WS$_2$ and MoS$_2$ monolayers \cite{cong2018intrinsic}. A combination of CVD-grown hBN and CVD-grown MoS$_2$ that can be stacked on top of each other has also been reported, where these stack showed superior optical quality to as-grown MoS$_2$ on SiO$_2$ \cite{wang2015all}. Also molecular beam epitaxy (MBE) is making progress for large scale growth of layered materials \cite{Zhang:2014a,xenogiannopoulou2015high,dau2018beyond}. Here for example large area films of MBE grown hBN can be used as a substrate to transfer TMD monolayer flakes. This heterostructure also showed overall spectrally narrow emission of the TMD ML \cite{doi:10.1063/1.5033554}.

As substrate and encapsulation techniques progress, ultimately the TMD material quality itself will need to be improved. For bulk material for exfoliation the flux growth techniques is reported to generate ML material with very low defect density \cite{doi:10.1021/acs.nanolett.9b00985,verzhbitskiy2019suppressed}. Here further improvement in the structural quality of CVD-grown layers needs to be investigated in the future. Additional decrease of the defect density in TMD samples grown by CVD can be achieved by optimization of the multiple thermodynamic and kinetic growth parameters including temperature, partial pressure of the components and their flow rates, as well as the TMD monolayer/substrate interaction. Moreover, the application of the gaseous precursors for transition metals and chalcogens in combination with the appropriate catalysts may provide an additional degree of freedom for optimizing the crystalline growth in the future.

\setcounter{figure}{0}

\section{Supporting Information}

\subsection{Experimental Methods}

\subsubsection{ High-resolution transmission electron microscopy - HRTEM}
The high-resolution transmission electron microscope (HRTEM) images were acquired with the C$_c$/C$_s$-corrected Sub-Angstrom Low-Voltage Electron microscope (SALVE). A voltage of 60~kV was used as we undercut the knock-on threshold for the Sulfur atoms \cite{PhysRevLett.109.035503} but still achieve sub-Angstrom resolution \cite{linck2016chromatic} with typical dose rates of about 10$^5$e$^-$/nm$^2$s. The values for the chromatic aberration Cc and the spherical aberration Cs were between -10~$\mu$m to -20~$\mu$m. HRTEM images of the CVD grown MoS$_2$ were acquired with bright atom contrast and recorded on a 4k$\times$4k camera with exposure times of 1~s. Exfoliated MoS$_2$ was also imaged with bright atom contrast and 1~s exposure time but recorded on a 2k$\times$2k GIF camera.

Bulk MoS$_2$, obtained from HQ Graphene, was exfoliated via adhesive tape on SiO$_2$ substrates. Using an optical microscope, monolayers were identified due to their contrast \cite{benameur2011visibility} on SiO$_2$. After locating MoS$_2$ monolayers, they were transferred to Quantifoil TEM grids R 1.2/1.3 by placing the grid with a drop of isopropyl alcohol on top. Due to the evaporation of the isopropyl alcohol, the grid comes into contact with the flake. In the next step, the SiO$_2$ is etched away with KOH and as a result the MoS$_2$ flake is released onto the TEM grid. To reduce residues after the transfer, the sample is cleaned with double distilled water.

\begin{figure*}
\includegraphics[width=0.85\linewidth]{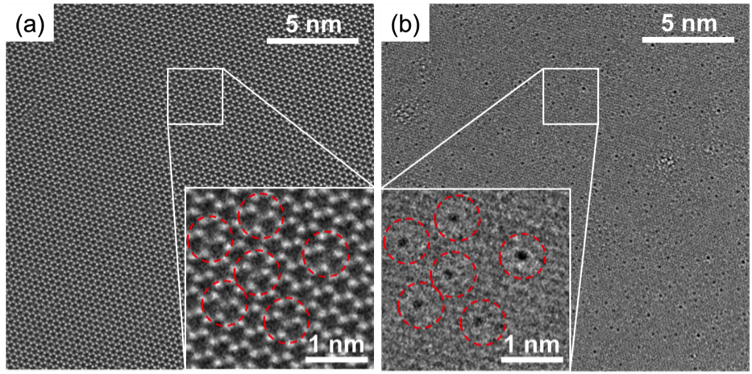}
\renewcommand{\thefigure}{S\arabic{figure}}
\caption{\label{fig:figS1} 60 kV HRTEM image of CVD-grown MoS$_2$. (a) shows the raw image of MoS$_2$. The area within the white square is magnified in the lower right. Red circles mark the vacancies which are difficult to see even in the magnified image. Thus, Fourier-filtering was applied to remove the frequencies of the MoS$_2$ lattice which is shown in (b). In (b), the same area like in (a) is magnified. Due to the Fourier-filtering the vacancies are better visible (black dots, surrounded by red circles).}
\end{figure*}

Image processing was performed to increase the visibility of the vacancies. Fig.~\ref{fig:figS1}a shows a raw image of MoS$_2$. In the magnified panel, vacancies are marked with red circles. For better visualization, a Fourier-filter to remove the frequencies of the MoS$_2$ lattice was applied. The result is shown in Fig.~\ref{fig:figS1}b. Small black dots in the lattice indicate a vacancy. In (b), the same area as in (a) is magnified and the vacancies are again marked with red circles. 

\subsubsection{Error Evaluation for defect concentration determination}
\begin{figure*}
\includegraphics[width=0.85\linewidth]{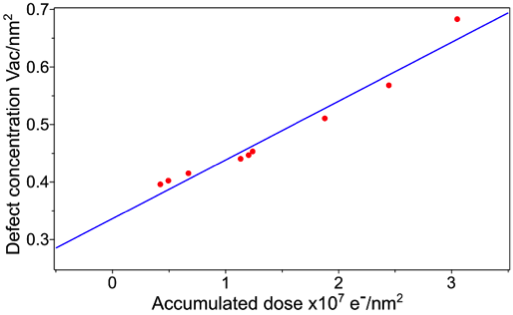}
\renewcommand{\thefigure}{S\arabic{figure}}
\caption{\label{fig:figS2} Measurements (red dots) of the vacancy (V) evolution per area depending on the accumulated doses $\phi$ in exfoliated MoS$_2$. Based on the slope of the linear fit (blue line), the error calculation was carried out. The starting vacancy concentration is probably slightly lower than the measured concentration because the evaluated area is always a few seconds exposed to the electron beam before the first image can be acquired. }
\end{figure*}

For the error evaluation, the evolution of the defect concentration was analysed (cf. Fig.~\ref{fig:figS2}). Furthermore, a linear behaviour for low defect concentrations (displaced S-atoms $< 5\%$) is assumed. Based on the slope for the linear defect evolution, the error of the defect concentration $\Delta C$ is determined as :

\begin{equation}
\Delta C = \sqrt{ \left(\frac{\Delta V}{A \cdot \phi}\right)^2+\left( \frac{V \cdot \Delta A}{A^2 \cdot \phi} \right)^2 + \left( \frac{V \cdot \Delta \phi}{A \cdot \phi^2} \right)^2}
\end{equation}

For the confidence intervals, we took $\Delta V = \sqrt{V}$ for the vacancies and $\Delta A = \sqrt{A}$ for the evaluated area. For the total accumulated dose, a huge confidence interval is assumed because of the uncertainties of the previous electron beam irradiation before the first image was recorded. Here, we assumed a preceding irradiation time of t = 10~s, so that the confidence interval for the accumulated dose, depending on the dose rate $\phi$, becomes $\Delta \phi = 10~s \cdot \phi$.

\subsubsection{Photoluminescence and differential reflectivity}
The optical spectroscopy experiments are carried out in a confocal microscope built in a vibration free, closed cycle cryostat from Attocube with variable temperature in the range T= 4 to 300~K. The excitation/detection spot diameter is below 1$\mu$m. Reflectivity measurements were performed with a power-stabilized white halogen light source. The optical signal is dispersed in a spectrometer and detected with a Si-CCD camera. For photoluminescence (PL) measurements we use either 633~nm or 532 nm laser excitation wavelength.

\begin{figure*}
\includegraphics[width=0.65\linewidth]{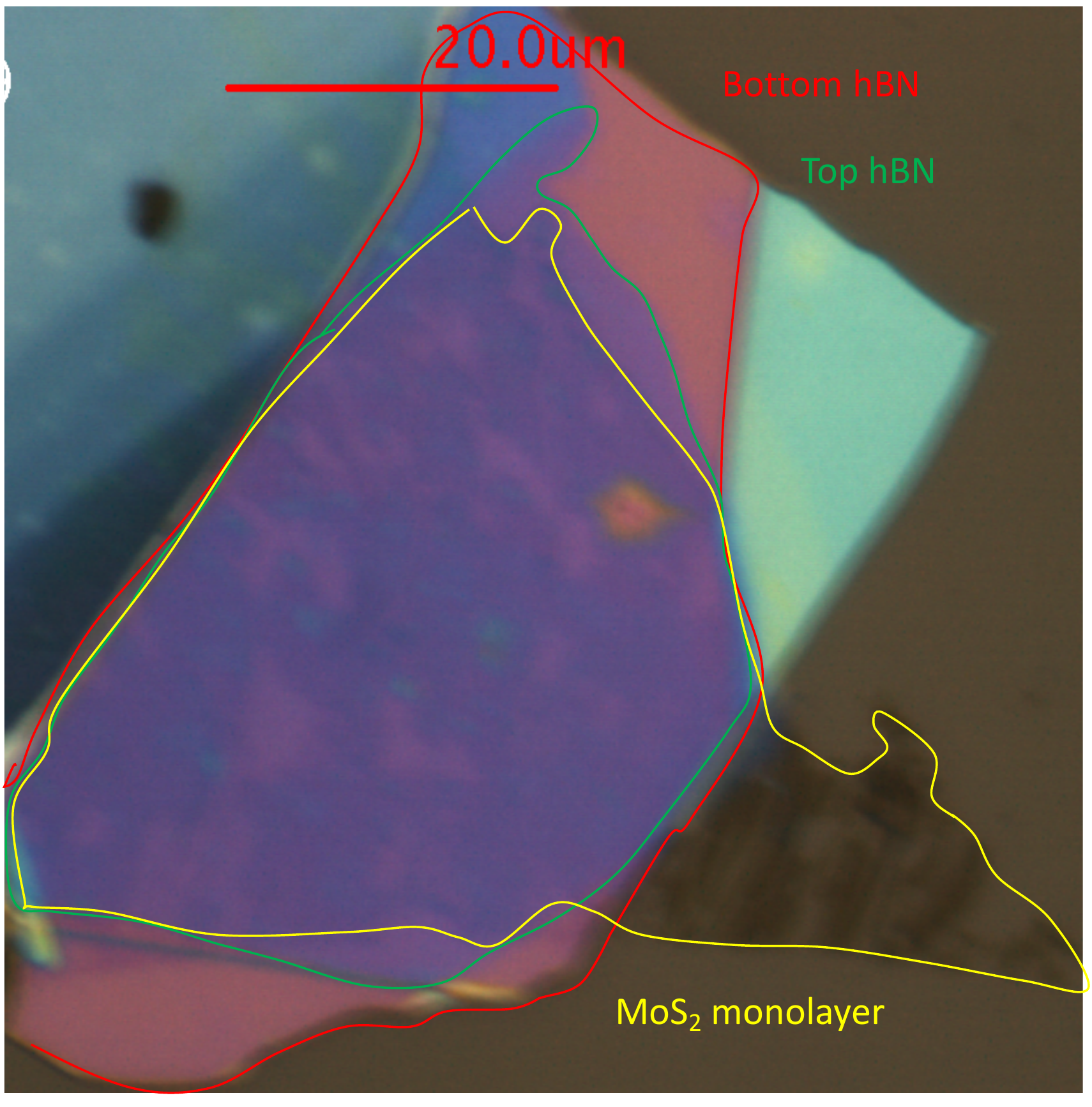}
\renewcommand{\thefigure}{S\arabic{figure}}
\caption{\label{fig:figS3} Sample 2. Optical microscope image of the CVD-grown MoS$_2$ monolayer (outline marked in yellow) sandwiched between an hBN bottom flake (marked in red) and an hBN top flake (marked in green). As the MoS$_2$ monolayer surface area is larger than the hBN flake area, there is also a ML part, directly in contact with the SiO$_2$ substrate.  }
\end{figure*}

In Fig.~\ref{fig:figS3} we show an optical microscope image of sample 2, the CVD-grown MoS$_2$ ML manually encapsulated in hBN using dry stamping \cite{Gomez:2014a}. Several samples have been fabricated using this procedure showing comparable, high optical quality.

\begin{figure*}
\includegraphics[width=0.65\linewidth]{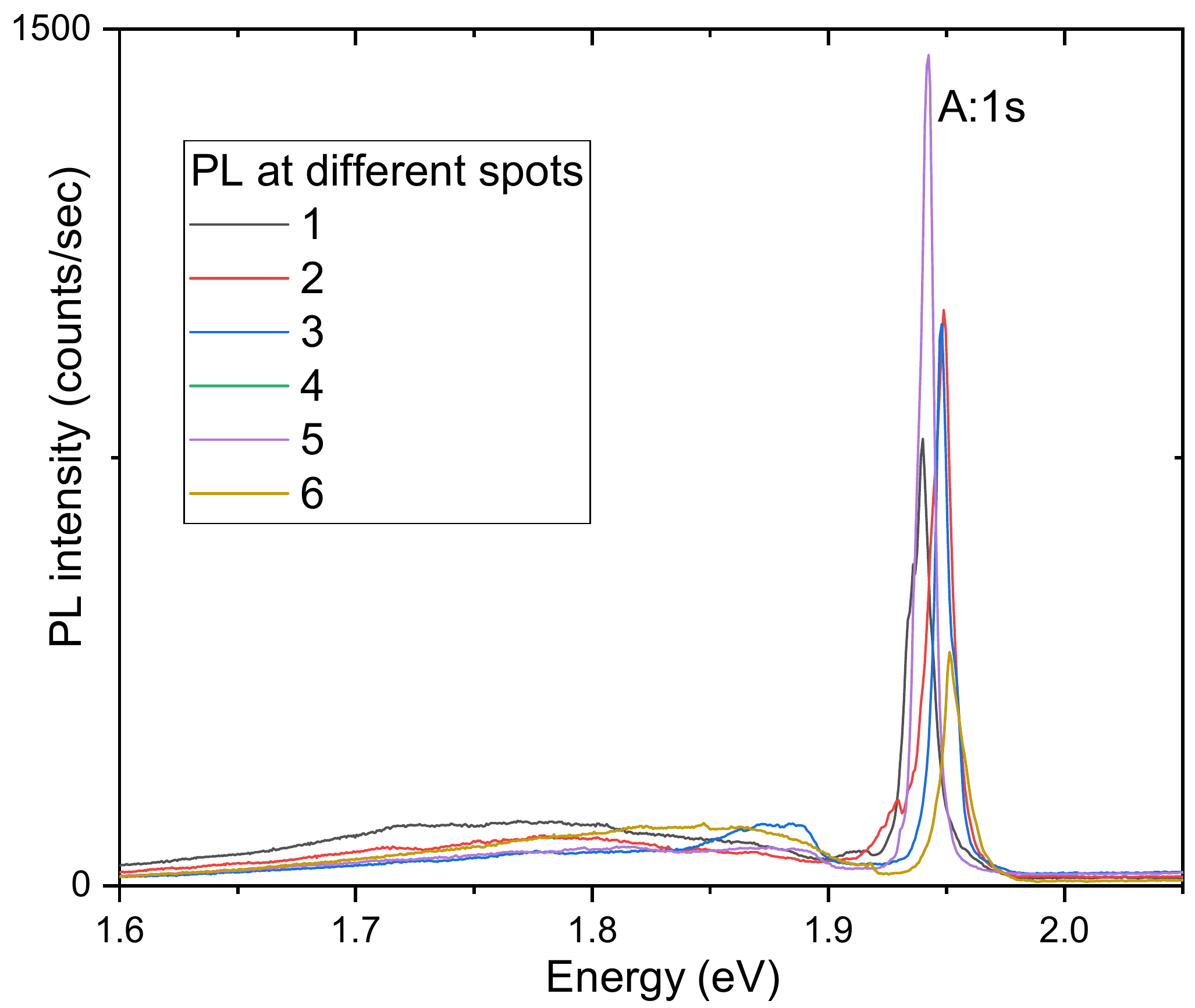}
\renewcommand{\thefigure}{S\arabic{figure}}
\caption{\label{fig:figS4} Sample 2. Photoluminescence emission at T=4~K from sample 2. Spectrally narrow emission (FWHM $\approx$5~meV) from the A-exciton (A:1$s$) is dominating independent of the exact sample area investigated. So for different positions of the detection spot on the sample the characteristics are very similar : weak defect emission at energies below 1.9~eV and strong exciton emission at around 1.95~eV from the CVD~grown MoS$_2$ monolayer. }
\end{figure*}

In Fig.~\ref{fig:figS4} we investigate the spatial inhomogeneity of the PL emission. Although there are small variations in energy (meV scale) our sample 2 shows narrow excitonic PL emission wand weak defect related emission for all detection spot positions.\\

 \textbf{Acknowledgements}  \\
S.S., A.G. and T.L. contributed equally to this work. M.Bs. permanent address : attocube systems AG, Eglfinger Weg 2, 85540 Haar bei M\"unchen. We thank Emmanuel Courtade for technical assistance and Sef Tongay for stimulating discussions.
Toulouse acknowledges funding from ANR 2D-vdW-Spin, ANR VallEx, ANR MagicValley, ITN Spin-NANO Marie Sklodowska Curie Grant Agreement No. 676108, ITN 4PHOTON No. 721394 and the Institut Universitaire de France. Growth of hexagonal boron nitride crystals was supported by the Elemental Strategy Initiative conducted by MEXT, Japan, and CREST (JPMJCR15F3), JST. This project has also received funding from the joint European Union’s Horizon 2020 and DFG research and innovation programme FLAG-ERA under a grant TU149/9-1, DFG Collaborative Research Center SFB 1375 “NOA” Project B2 and the Thüringer MWWDG via FGR 0088 “2D-Sens”.


\end{document}